\documentclass[twocolumn,secnumarabic,amssymb, nobibnotes, aps, prd]{revtex4}
\usepackage{graphicx}

\begin{document}
\title{Can dc voltage proportional to the persistent current be observed on segment of asymmetric mesoscopic ring?}
\author{S.V. Dubonos, V.L. Gurtovoi, A.V. Nikulov, and V.A. Tulin}
\affiliation{Institute of Microelectronics Technology and High Purity Materials, Russian Academy of Sciences, 142432 Chernogolovka, Moscow District, RUSSIA.} 
\begin{abstract} In order to clear up a question on possibility of a dc voltage proportional to the persistent current in normal metal and semiconductor mesoscopic loops conjectural causes of this phenomenon observed in superconductor loops is investigated.
 \end{abstract}

\maketitle

\narrowtext

\section*{Introduction}

I.O.Kulik had shown in theory 1970 that the persistent current can be observed both in normal state of superconductor at $T > T_{c}$ [1] and in non-superconductor mesoscopic structures [2]. An experimental evidence of the persistent current in the fluctuation region of superconductor at $T \geq  T_{c}$ was obtained as far back as 1962 [3], before the theory [1]. According to the universally recognized explanation [4] the resistance oscillation in magnetic field $\Delta R(\Phi/\Phi_{0}) \propto \Delta T_{c}(\Phi/\Phi_{0}) $ of superconducting cylinder [3] or loop [5] is a consequence of the persistent current oscillations $I_{p}(\Phi/\Phi_{0})$. The Little-Parks oscillations of the loop critical temperature $\Delta T_{c}(\Phi/\Phi_{0}) $ [3-6] are observed since the velocity circulation $\oint_{l} dl v_{s} = (1/m)(n2\pi \hbar  + 2e\Phi) = (2\pi \hbar /m)(n  + \Phi/\Phi_{0}) $ can not be equal zero at the magnetic flux $\Phi $ inside the loop (or cylinder) not divisible by the flux quantum $\Phi_{0} = \pi \hbar /2e$, $\Phi \neq n\Phi_{0}$, because of the quantization $\oint_{l} dl p = \oint_{l} dl  (mv_{s} + 2eA) = m \oint_{l} dl v_{s} + 2e\Phi = n2\pi \hbar $.

The persistent current has measurable value in real superconducting loop even at $T \geq  T_{c}$. For example, according to the experimental data presented in [5] the amplitude of the persistent current, $I_{p,A} = |I_{p}|$ at $\Phi = (n+0.5)\Phi_{0}$, of a square loop with a size $1\times 1 \ \mu m$ and wire section $s = w \times d = 0.15 \times 0.025 \ \mu m \approx 0.004 \ \mu m^{2}$ is equal approximately $I_{p,A} \approx 0.2 \ \mu A$ at the midpoint of the resistive superconducting transition $T \approx  T_{c}$. This value is much larger than a current of two electrons $I_{2e} = 2ev_{s}/l \approx 10^{-11} \ A$ at the velocity $|v_{s}| = (2\pi \hbar /ml)(0.5) \approx 100 \ m/c$ permitted in this loop with the perimeter $l = 4 \ \mu m$ at $\Phi = (n+0.5)\Phi_{0}$ since superconducting pairs, as condensed bosons, have the same velocity and therefore $I_{p} = sj_{p} = s2e\overline{n_{s}}v_{s}$. The average fluctuation density of superconducting pairs $\overline{n_{s}} = I_{p}/s2ev_{s} = I_{p}/s2ev_{s} \approx 3 \ 10^{24} \ m^{-3}$ in the Al loop [5] at $T \approx  T_{c}$ corresponds to $\approx 10^{-5}$ of the total density of electron pairs in aluminum. The great number of pairs in the loop, $N_{s} = sl\overline{n_{s}} \approx 10^{5}$ at $T \approx  T_{c}$, explains why the persistent current can be observed even in the fluctuation region, in spite of the great size $l$ in comparison with the one of atom orbit: the energy difference between adjacent permitted states $E_{n+1} - E_{n} \approx N_{s}2\pi ^{2}\hbar ^{2}/ml^{2} \approx k_{B} 1000 \ K  \gg k_{B}T_{c}$ whereas for single electron $E_{n+1} - E_{n} \approx 2\pi ^{2}\hbar ^{2}/ml^{2} < k_{B} 0.1 \ K$ [7].

The persistent current is much weaker and it can be observed only at very low temperature in non-superconductor loop since $2\pi ^{2}\hbar ^{2}/ml^{2} < k_{B} 0.1 \ K$ for a real size $l$. Therefore first attempts to observe this quantum phenomenon in normal metal [8] and semiconductor [9] loops were made 20 years later its prediction [2] and 28 years later its observation in superconductor at $T \geq T_{c}$ [3]. The typical amplitude of the persistent current oscillations measured on Cu [8], Au [10,11], GaAlAs/GaAs [9,12-14], Ag [15] mesoscopic loops estimated to be of the order of $10^{-10} \ A$. In order to increase the magnetic response from this weak current systems with great number of loops, $10^{5}$ [12,14,15] and even $10^{7}$ [8] were used. Quite enough response could be observed in a system with such number of asymmetric loops if an analogy with conventional circular current is valid.

\section {Analogy with conventional circular current}
It is well known that a potential difference $V = (R_{ls} - R_{l}l_{s}/l)I$ is observed on a segment $l_{s}$ (with a resistance $R_{ls}$) of an asymmetric conventional metal loop $l$ (with a resistance $R_{l}$) when a circular current $I = \oint_{l} dl E/R_{l}$ is induced by the Faraday's voltage $\oint_{l} dl E = -d\Phi/dt$. It was found experimentally [16,17] that there is an analogy between this conventional current and the persisent current in superconductor loop in spite of the Faraday's voltage absence $d\Phi/dt = 0$ in the second case. The observations of the quantum oscillations of the dc potential difference $V_{dc}(\Phi/\Phi_{0}) \propto I_{p}(\Phi/\Phi_{0})$ on asymmetric aluminum rings at $T \approx  T_{c}$ [16] and $T <  T_{c}$ [17] raise a question about possibility of like phenomenon in asymmetric normal metal and semiconductor mesoscopic loops with $I_{p}(\Phi/\Phi_{0})$. It was proved experimentally [17] that  the $V_{dc}(\Phi/\Phi_{0})$ amplitude increases with number of asymmetric superconductor loops connected in series. The measurable $V_{dc}(\Phi/\Phi_{0})$ oscillations with the amplitude up to $1 \ \mu V$ could be expected on a system of $10^{5}$ loops with $I_{p,A} \approx 10^{-10} \ A$ and resistance asymmetry $(R_{ls} - R_{l}l_{s}/l) \approx 0.1 \ \Omega $ if the analogy with conventional circular current is valid.

Possibility of this analogy for the persistent current in normal metal and semiconductor loops seems paradoxical. But the observations [3-6,8-17] of the dc circular current $I_{p} \neq 0$ in loops with non-zero resistance $R > 0$ and without the Faraday's voltage $d\Phi/dt = 0$ is already incomprehensible paradox. The statement [18] that the existence of a finite Ohmic resistance for a phase coherence sample is not paradoxical when one properly takes into account the influence of the measuring leads is disproved with simultaneous observations $I_{p} \neq 0$ and $R > 0$ of loops with leads made in numerous works [3-6,16,17]. The $V_{dc}(\Phi/\Phi_{0})$ oscillations were already observed in superconductor loops [16,17]. In order to clear up a question on possibility this phenomenon in normal metal and semiconductor loops its cause in superconductor loops should be investigated comprehensively.

\begin{figure}[b]
\includegraphics{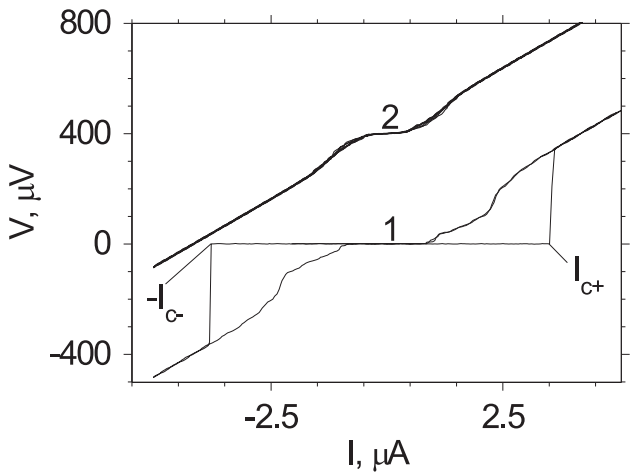}
\caption{\label{fig:epsart} The current-voltage curves of system of 18 asymmetric Al loops connected in series measured at different temperature 1) $T \approx 0.984T_{c}$ and 2) $T \approx 0.995T_{c}$.}
\end{figure}

\begin{figure}
\includegraphics{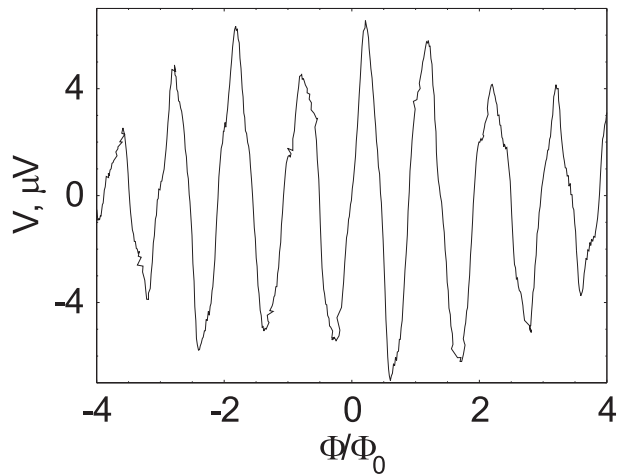}
\caption{\label{fig:epsart} The quantum oscillations of the dc potential difference $V(\Phi/\Phi_{0})$ induced by the ac current with frequency $f = 40 \ kHz$ and amplitude $I_{0} = 3 \ \mu A$  on system of 18 asymmetric Al loops near superconducting transition $T \approx 0.996T_{c}$ where the hysteresis of the current-voltage curves is absent, see Fig.1. }
\end{figure}

\section{Conjectural causes of the quantum oscillations $V_{dc}(\Phi/\Phi_{0})$ observed in superconducting loops}
We have shown in [19] that the rectification of ac current observed at $T < 0.99T_{c}$ is consequence of the critical current anisotropy of asymmetric superconducting loops. The current-voltage curves (CVC) in this temperature region have a hysteresis and sharp transition in the resistance state at $I_{c+}$ or $I_{c-}$, Fig.1. We have found that the value of the critical current depends on direction of its measurement $I_{c+} \neq I_{c-}$ and that the sign and value of this anisotropy $I_{c,an} = I_{c+} - I_{c-}$ are periodical function of magnetic flux $I_{c,an}(\Phi/\Phi_{0})$ [19]. Although the cause of the anisotropy $I_{c,an}(\Phi/\Phi_{0})$ is very strange [20] the CVC and the proportionality $V_{dc}(\Phi/\Phi_{0}) \propto -I_{c,an}(\Phi/\Phi_{0})$ [19] allow to explain the dc voltage $V_{dc}(\Phi/\Phi_{0})$ observed at $T < 0.99T_{c}$ as a result of the transition in the normal state of the whole loop at different values $I_{c+}$ and $I_{c-}$. This case is not useful for the clearing up of the question on the $V_{dc}(\Phi/\Phi_{0})$ possibility in normal metal and semiconductor mesoscopic loops.

More useful may be our results obtained at $T > 0.99T_{c}$ where the CVC are smooth and reversible, Fig.1. The voltage observed at $I_{p} \neq 0$ and $R < R_{n}$ can be reversible only if the loop is switched between superconducting states with different connectivity of wave function. Because of such switching induced by thermal fluctuations the persistent current $I_{p} \neq 0$ can be observed at $R > 0$ [7]. The break of the wave function describing superconducting state in loop segments is similar to electron scattering in mesoscopic loop with the persistent current. Therefore the observation of the quantum oscillations $V_{dc}(\Phi/\Phi_{0})$ in superconducting loop at $T > 0.99T_{c}$, Fig.2, allow to expect that such quantum oscillations can be observed also in normal metal and semiconductor mesoscopic loops. It is no coincidence that I.O. Kulik made the works [1,2] on superconductor and non-superconductor mesoscopic structures in the same year. The persistent current has the same cause in the both cases. Therefore the results of the investigations of the $V_{dc}(\Phi/\Phi_{0})$  oscillations in superconductor loop can give an information about possibility this phenomenon in normal metal and semiconductor loops.

\section*{Acknowledgement}
This work has been supported by a grant of the Program "Quantum Nanostructures" of the Presidium of RAS, grant "Quantum bit on base of micro- and nano-structures with metal conductivity" of the Program "Technology Basis of New Computing Methods" of ITCS department of RAS and a grant 04-02-17068 of the  Russian Foundation of Basic Research.

\end{document}